\documentclass{nature_millenium}

\usepackage{chapterbib}
\usepackage{times} 
\usepackage{mathptmx}
\usepackage[small,bf,format=plain]{caption}
\usepackage{sectsty}
\usepackage{amssymb}
\usepackage{graphicx}
\usepackage{enumerate}
\usepackage{bm}
\usepackage{authblk}

\typearea{12}
\paragraphfont{\small}
\subsectionfont{\normalsize}

\def \eV          {\rm eV}
\def \pc          {\rm pc}
\def \kpc         {\rm kpc}
\def \Mpc         {\rm Mpc}
\def \msun        {M_\odot}
\def \psiDM       {${\psi}$DM}
\def \LambdapsiDM {${\Lambda\psi}$DM}

\newcommand{\appropto}{\mathrel{\vcenter{
  \offinterlineskip\halign{\hfil$##$\cr
  \propto\cr\noalign{\kern2pt}\sim\cr\noalign{\kern-2pt}}}}}

\begin{document}

\title{\vspace*{-1cm}{\Large Cosmic Structure as the Quantum Interference of
a Coherent Dark Wave}}

\author{{\small\sffamily\textbf{
Hsi-Yu Schive$^{1}$,
Tzihong Chiueh$^{1,2\ast}$ \&
Tom Broadhurst$^{3,4}$}}}

\affil[]{\normalsize\textbf{\textit{chiuehth@phys.ntu.edu.tw}}\vspace{-0.2cm}}

\date{}

\twocolumn

\baselineskip14pt 
\setlength{\parskip}{4pt}
\setlength{\parindent}{18pt}%
\setlength{\footskip}{25pt}
\setlength{\textheight}{670pt}
\setlength{\oddsidemargin}{-8pt}
\setlength{\topmargin}{-41pt}
\setlength{\headsep}{18pt}
\setlength{\textwidth}{469pt}
\setlength{\marginparwidth}{42pt}
\setlength{\marginparpush}{5pt}

\addtolength{\topmargin}{-0.6cm}

\setcounter{figure}{0}
\renewcommand{\figurename}{\textbf{Figure}}
\renewcommand{\thefigure}{\textbf{\arabic{figure}}}

{\bf\small
\twocolumn[\begin{@twocolumnfalse}
\maketitle

\begin{abstract}
The conventional cold, particle interpretation of dark matter (CDM)
still lacks laboratory support and struggles with the basic properties
of common dwarf galaxies, which have surprisingly uniform central masses
and shallow density profiles\cite{Gilmore2007,Strigari2008,WP2011,AAE2013}.
In contrast, galaxies predicted by CDM extend to much lower masses, with
steeper, singular profiles\cite{Kauffmann1993,Klypin1999,Moore1999}.
This tension motivates cold, wavelike dark matter ({\psiDM}) composed of a
non-relativistic Bose-Einstein condensate, so the uncertainty principle counters
gravity below a Jeans scale\cite{Peebles2000,Hu2000}. Here we achieve
the first cosmological simulations of this quantum state at unprecedentedly
high resolution capable of resolving dwarf galaxies, with only one free
parameter, $\bf{m_B}$, the boson mass. We demonstrate the large scale
structure of this {\psiDM} simulation is indistinguishable from CDM, as
desired, but differs radically inside galaxies. Connected filaments and
collapsed haloes form a large interference network, with gravitationally
self-bound solitonic cores inside every galaxy surrounded by extended haloes of
fluctuating density granules. These results allow us to determine
$\bf{m_B=(8.1^{+1.6}_{-1.7})\times 10^{-23}~eV}$ using stellar
phase-space distributions in dwarf spheroidal galaxies. Denser, more
massive solitons are predicted for Milky Way sized galaxies, providing
a substantial seed to help explain early spheroid formation. Suppression of
small structures means the onset of galaxy formation for {\psiDM} is
substantially delayed relative to CDM, appearing at $\bf{z\lesssim 13}$
in our simulations.
\end{abstract}

\end{@twocolumnfalse}]
}

\small

\footnotetext[1]{\footnotesize Dept. of Physics, National Taiwan Univ.,
                               Taipei 10617, Taiwan}
\footnotetext[2]{\footnotesize Center for Theoretical Sciences, National
                               Taiwan Univ., Taipei 10617, Taiwan}
\footnotetext[3]{\footnotesize Dept. of Theoretical Physics, Univ. of the
                               Basque Country UPV/EHU, E-48080 Bilbao, Spain}
\footnotetext[4]{\footnotesize Ikerbasque, Basque Foundation for Science,
                               E-48011 Bilbao, Spain}

Thermally produced particle candidates for the dark matter are still
without laboratory support, including those favoured by
super-symmetric theories in the TeV range.
Non-thermal bosonic fields,
particularly scalar fields, provide another well motivated class of
dark matter, formed in a non-relativistic, low-momentum state as a
cold Bose-Einstein condensate (BEC), and increasingly motivated by
extensions of the Standard Model of particle physics and to the mechanism
driving the universal expansion\cite{Peebles1988}.
The field in this context can be described by a coherent wave function $\psi$
with an interference pattern determining the distribution of dark
matter, which we term {\psiDM}. Axion is a long-standing CDM
candidate of this form, and higher dimensional theories motivate an
``axiverse'', where a discrete mass spectrum of axion-like particles spans
many decades, possibly affecting cosmic structure\cite{Arvanitaki2010}.

\begin{figure*}[t]
\centering
\includegraphics[width=7.9cm]{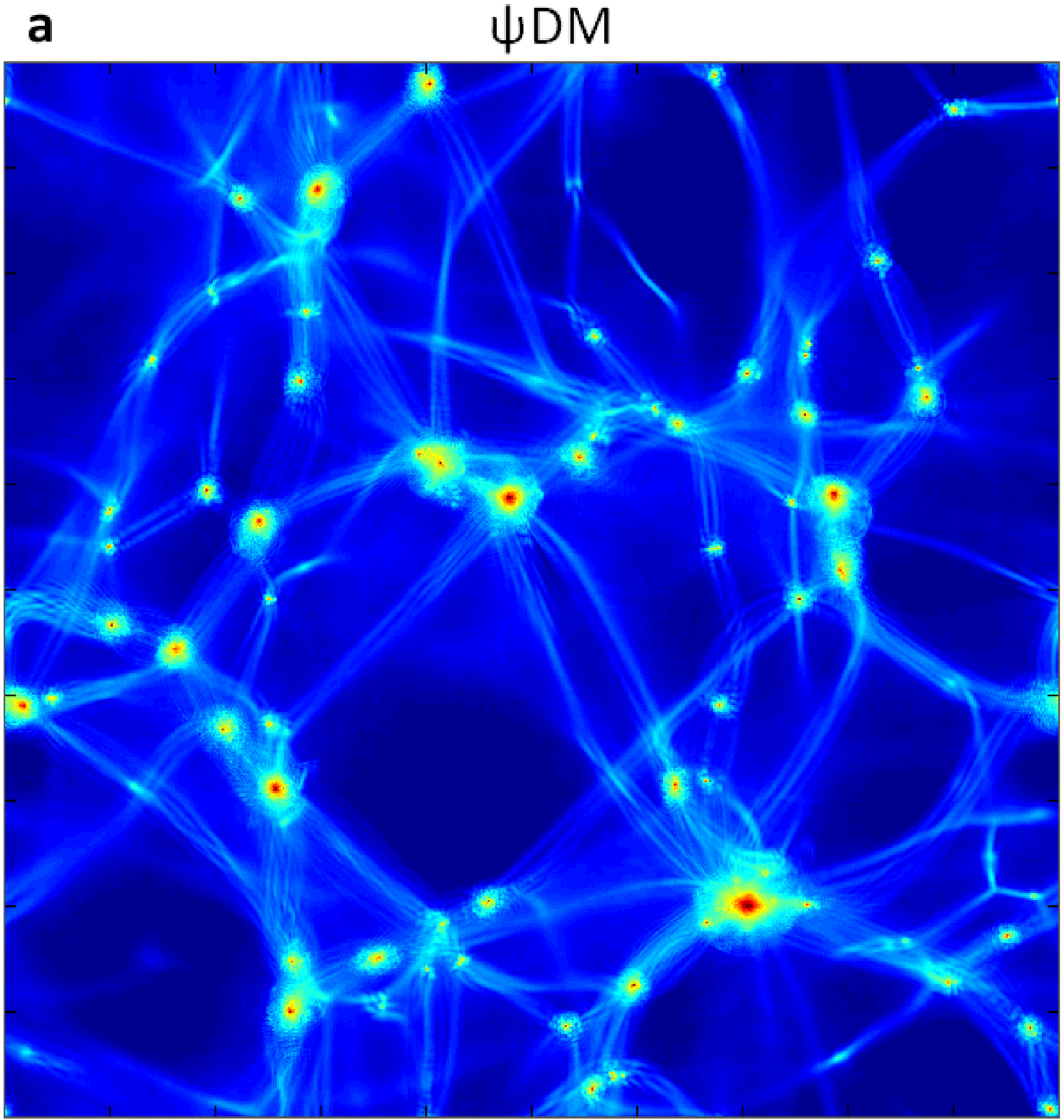}
\includegraphics[width=7.9cm]{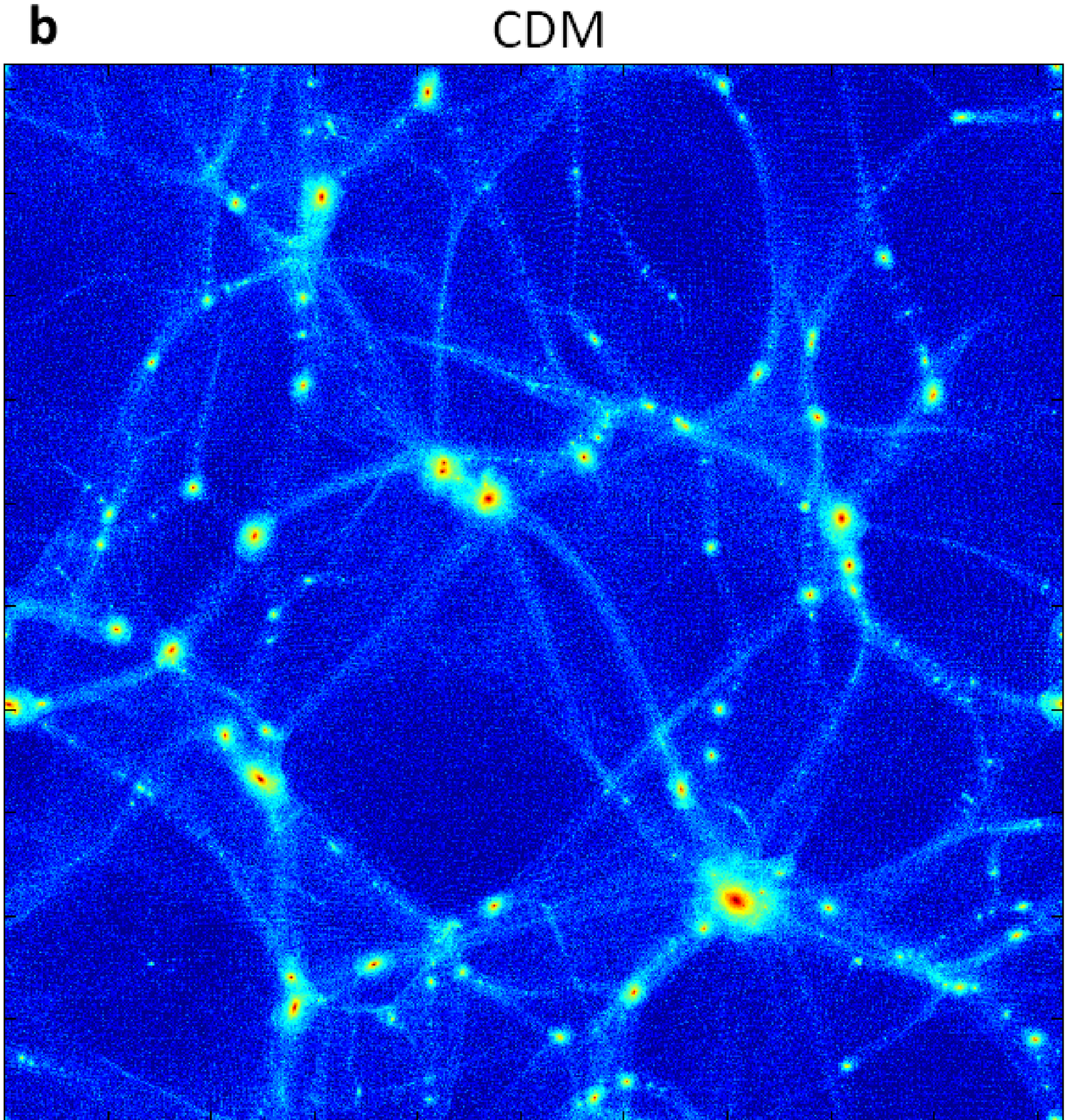}
\caption{\textbf{Comparison of cosmological large-scale structures formed by
standard CDM and by wavelike dark matter, {\psiDM}.}
Panel (\textbf{a}) shows the structure created by evolving a single coherent
wave function for {\LambdapsiDM} calculated on AMR grids.
Panel (\textbf{b}) is the structure simulated with a standard $\Lambda$CDM N-body code
GADGET-2\cite{GADGET2} for the same cosmological parameters, with the high-k
modes of the linear power spectrum intentionally suppressed in
a way similar to the {\psiDM} model to
highlight the comparison of large-scale features. This comparison
clearly demonstrates that the large scale distribution of filaments
and voids is indistinguishable between these two completely different
calculations, as desired given the success of $\Lambda$CDM in
describing the observed large scale structure. {\psiDM} arises from the
low momentum state of the condensate so that it is equivalent to
collisionless CDM well above the Jeans scale.}
\label{fig:CDM_vs_BEC}
\end{figure*}

The distribution of {\psiDM} mimics particle CDM on
large scales\cite{Widrow1993,Sikivie2009,WC2009}, and hence
distinguishing between CDM and cold, wavelike {\psiDM} is best made on small scales
due to the additional quantum stress\cite{Peebles2000,Hu2000,WC2009}.
Dwarf spheroidal (dSph) galaxies are the smallest and most common class of
galaxy with internal motions dominated by dark matter. Their basic
properties are very hard to explain with standard CDM, including
the surprising uniformity of their central masses, $M(<300~pc)\simeq
10^7~\msun$, and shallow density profiles\cite{Gilmore2007,Strigari2008,WP2011,AAE2013}.
In contrast, galaxies predicted by
CDM extend to much lower masses, well below the observed dwarf
galaxies, with steeper, singular mass
profiles\cite{Kauffmann1993,Klypin1999,Moore1999}.
Adjustments to standard CDM addressing
these difficulties consider particle collisions\cite{SS2000}, or warm dark
matter (WDM)\cite{Bode2001}. WDM can be tuned to suppress
small scale structures, but does not provide large enough flat
cores\cite{Strigari2006,Maccio2012}. Collisional CDM can be adjusted
to generate flat cores, but cannot suppress low mass galaxies without
resorting to other baryonic physics\cite{Rocha2013}.
Better agreement is expected for {\psiDM} because the uncertainty principle
counters gravity below a Jeans scale, simultaneously suppressing small scale
structures and limiting the central density of collapsed
haloes\cite{Peebles2000,Hu2000}.

\begin{figure*}[t]
\centering
\vspace*{0.2cm}
\includegraphics[width=17cm]{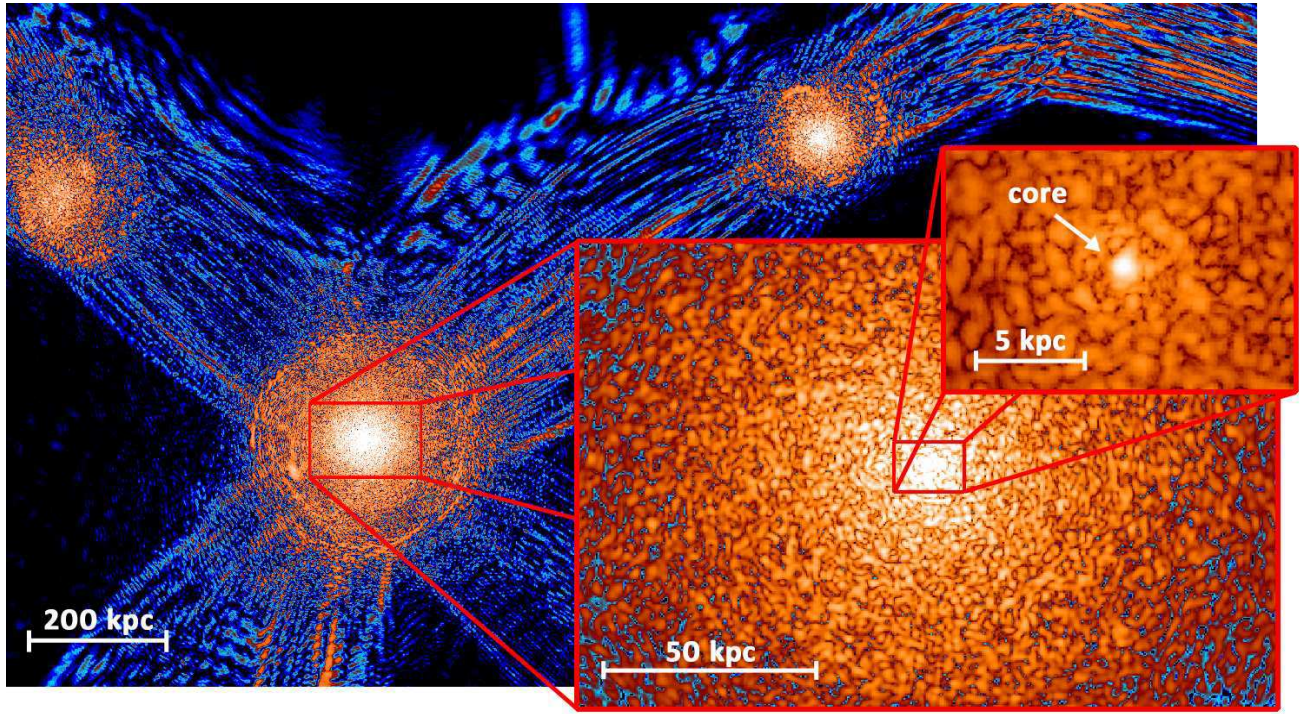}
\caption{\textbf{A slice of density field of {\psiDM} simulation on various
scales at $\bm{z=0.1}$.}
This scaled sequence (each of thickness 60 pc) shows how quantum interference
patterns can be clearly seen everywhere from the large-scale
filaments, tangential fringes near the virial boundaries, to the
granular structure inside the haloes. Distinct solitonic cores with radius
$\sim 0.3-1.6~\kpc$ are found within each collapsed halo.
The density shown here spans over nine orders of magnitude,
from $10^{-1}$ to $10^8$ (normalized to the cosmic mean density).
The color map scales logarithmically, with cyan
corresponding to density $\lesssim 10$.}
\label{fig:DensSlice}
\end{figure*}

Detailed examination of structure formation with {\psiDM} is therefore
highly desirable, but, unlike the extensive N-body investigation of
standard CDM, no sufficiently high resolution simulations of {\psiDM} have
been attempted. The wave mechanics of {\psiDM}
can be described by Schr\"{o}dinger's equation, coupled to gravity via
Poisson's equation\cite{Widrow1993} with negligible microscopic
self-interaction. The dynamics here differs from collisionless
particle CDM by a new form of stress tensor from quantum uncertainty,
giving rise to a comoving Jeans length, $\lambda_J\propto
(1+z)^{1/4}m_B^{-1/2}$, during the matter-dominated
epoch\cite{WC2009}. The insensitivity of $\lambda_J$ to redshift, $z$,
generates a sharp cutoff mass below which structures are
suppressed. Cosmological simulations in this context turn out to be
much more challenging than standard N-body simulations as the highest
frequency oscillations, $\omega$, given approximately by the matter
wave dispersion relation, $\omega \propto m_B^{-1}\lambda^{-2}$, occur
on the smallest scales, requiring very fine temporal resolution even
for moderate spatial resolution (see Supplementary Fig. S1). In this
work, we optimise an adaptive-mesh-refinement (AMR) scheme, with
graphic processing unit acceleration, improving performance by almost
two orders of magnitude\cite{GAMER2010} (see Supplementary Section 1
for details).

Fig. 1 demonstrates that despite the completely different calculations
employed, the pattern of filaments and voids generated by a
conventional N-body particle $\Lambda$CDM simulation is remarkably
indistinguishable from the wavelike {\LambdapsiDM} for the same linear
power spectrum (see Supplementary Fig. S2). Here $\Lambda$ represents
the cosmological constant. This agreement is desirable given the
success of standard $\Lambda$CDM in describing the statistics of large
scale structure. To examine the wave nature that distinguishes
{\psiDM} from CDM on small scales, we resimulate with a very high
maximum resolution of $60~\pc$ for a 2 Mpc comoving box, so that the
densest objects formed of $\gtrsim 300~\pc$ size are well resolved
with $\sim 10^3$ grids. A slice through this box is shown in Fig. 2,
revealing fine interference fringes defining long filaments, with
tangential fringes near the boundaries of virialized objects, where
the de Broglie wavelengths depend on the local velocity of matter. An
unexpected feature of our {\psiDM} simulations is the generation of
prominent dense coherent standing waves of dark matter in the center
of every gravitational bound object, forming a flat core with a sharp
boundary (Figs. 2 and 3). These dark matter cores grow as material is
accreted and are surrounded by virialized haloes of material with
fine-scale, large-amplitude cellular interference, which continuously
fluctuates in density and velocity generating quantum
and turbulent pressure support against gravity.

\begin{figure}[t]
\centering
\vspace*{0.2cm}
\includegraphics[width=7.9cm]{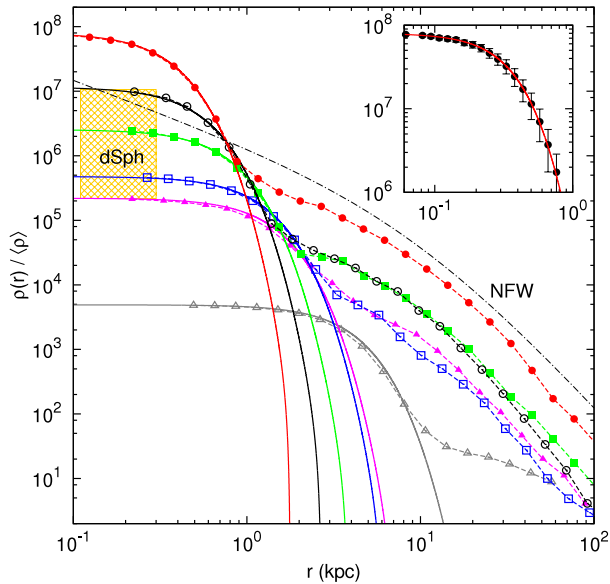}
\caption{\textbf{Radial density profiles of haloes formed in the {\psiDM} model.}
Dashed lines with various symbols show six examples of the halo
profiles normalized to the cosmic mean density. All haloes are found
to possess a distinct inner core fitted extremely well by the
soliton solution (solid lines).
A detailed soliton fit for the largest halo is inset, where the error
is the root-mean-square scatter of density in each radial bin.
An NFW profile representing standard CDM is also shown
for comparison (black dot-dashed line, with a very large scale radius of $10~\kpc$),
which fits well the profiles outside the cores.
The yellow hatched area indicates the $\rho_{300}$ of the dSph satellites
around Milky Way\cite{Strigari2008,AE2011},
which is consistent with the majority of galaxy haloes formed
in the {\psiDM} simulations.}
\label{fig:DensProfile}
\end{figure}

The central density profiles of all our collapsed cores fit well with the
stable soliton solution of the Schr\"{o}dinger-Poisson equation, as
shown in Fig. 3 (see also Supplementary Section 2 and Fig. S3).
On the other hand, except for the
lightest halo which has just formed and is not yet virialized, the
outer profiles of other haloes possess a steepening logarithmic slope,
similar to the Navarro-Frenk-White (NFW) profile\cite{NFW} of standard
CDM. These solitonic cores, which are gravitationally self-bound and appear
as additional mass clumps superposed on the NFW profile, are clearly distinct
from the cores formed by WDM and collisional CDM which truncate the NFW cuspy
inner profile at lower values and require an external halo for confinement.
The radius of the soliton scales inversely with mass, such that
the widest cores are the least massive and are hosted by the least massive
galaxies. Eighty percent of the haloes in the simulation have an
average density within $300~\pc$ (defined as $\rho_{300}$) in the range
$5.3\times10^{-3}-6.1\times10^{-1}~\msun/\pc^3$, consistent with Milky Way
satellites\cite{Strigari2008,AE2011}, and objects like these are resilient
to close interaction with massive galaxies. By contrast,
the very lowest mass objects in our simulation have $\rho_{300}\sim 4.0\times10^{-4}~\msun/\pc^3$, but
exist only briefly as they are vulnerable to tidal disruption by large
galaxies in our simulations.

The prominent solitonic cores uncovered in our simulations provide an
opportunity to estimate the boson mass, $m_B$, by comparison with
observations, particularly for dSph galaxies where dark
matter dominates. The local Fornax dSph galaxy is the best studied
case with thousands of stellar velocity measurements, allowing a
detailed comparison with our soliton mass profile. We perform a
Jeans analysis for the dominant intermediate metallicity stellar
population, which exhibits a nearly uniform projected velocity
dispersion ($\sigma_{||}$)\cite{AE2012a}. We
simultaneously reproduce well the radial distribution of the stars\cite{AE2012a}
(Fig. 4a) and their velocity dispersion with
negligible velocity anisotropy, with $m_B=(8.1^{+1.6}_{-1.7})\times10^{-23}~\eV$
and core radius $r_c=0.92^{+0.15}_{-0.11}~\kpc$ (see Supplementary Fig. S4).
The corresponding core mass $M(r\le r_c)$ is $\simeq 9.1\times10^7~\msun$,
which is hosted by a halo with virial mass $\simeq 4\times10^9\msun$
in the simulations. These results are similar to other estimates for
Fornax\cite{Wolf2010,Cole2012,AAE2013}
(Fig. 4b) and consistent with other dSph
galaxies derived by a variety of means\cite{WP2011,Lora2012,Wolf2010}
(see Supplementary Section 3 for details).

\begin{figure*}[t]
\centering
\includegraphics[width=7.8cm]{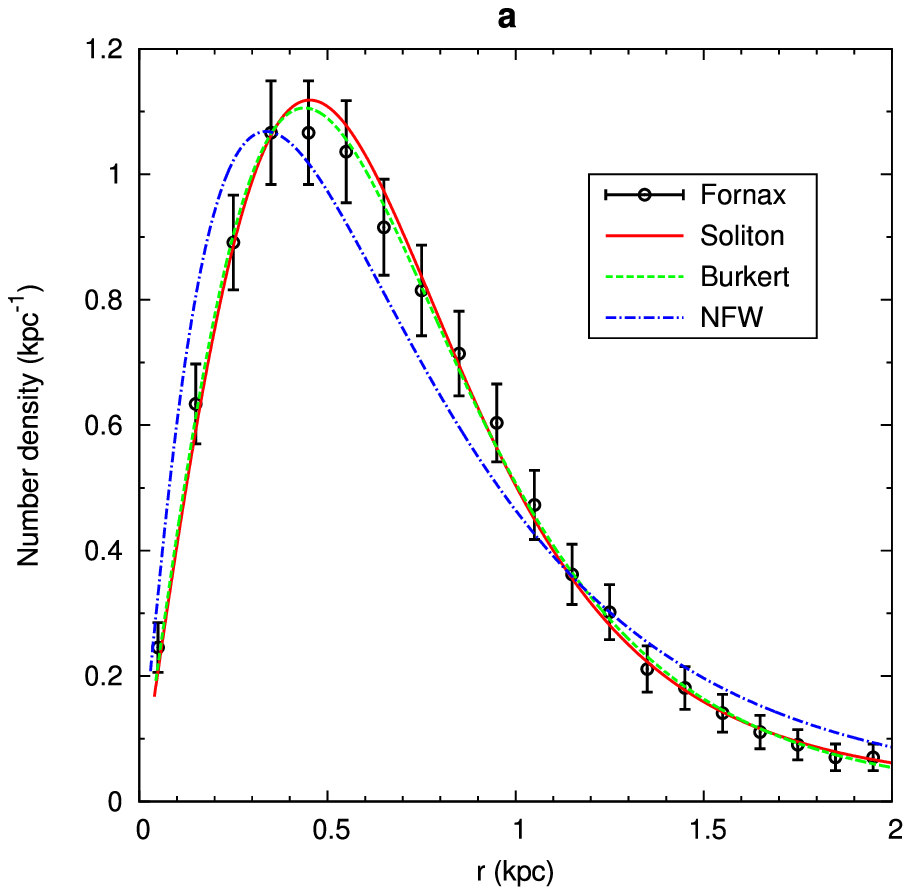}
\includegraphics[width=8.0cm]{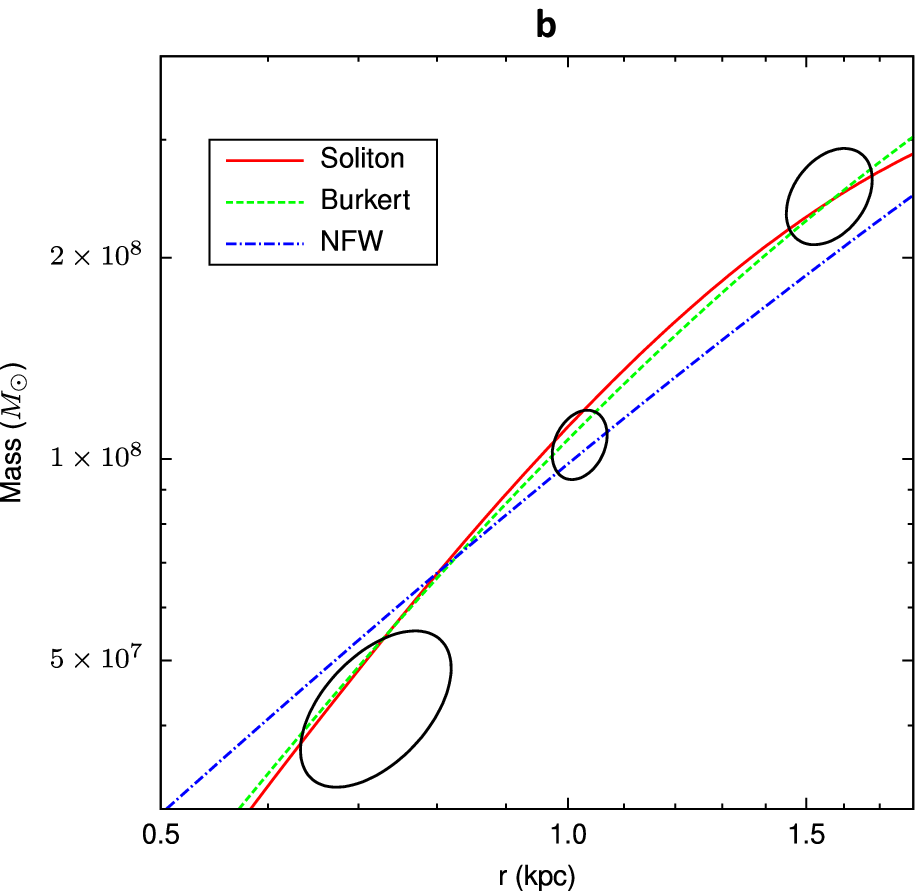}
\caption{\textbf{Modeling the Fornax dSph galaxy with the soliton profile.}
Panel (\textbf{a}) shows the normalized
stellar number density of the intermediate metallicity subpopulation\cite{AE2012a}
(symbols with 1-$\sigma$ error bars) and the best-fit soliton solution
(red solid line) with $m_B=8.1\times 10^{-23}~\eV$, $r_c=0.92~\kpc$,
and $\sigma_{||}=11.3~{\rm km/s}$. Also shown are the best-fit
empirical formula of Burkert\cite{Burkert1995} (green dashed line) and the NFW profile (blue dot-dashed line)
representing standard CDM. The scale radius of NFW is
restricted to be no larger than $3.0~\kpc$ during the fit to exclude
unreasonably small concentration parameters.
Panel (\textbf{b}) shows the 1-$\sigma$ contours
of the total enclosed mass estimated from each of the three
subpopulations\cite{AAE2013}, overplotted with the model curves using
the same best-fit parameters adopted in panel (\textbf{a}).
Clearly, in both panels the soliton profile of {\psiDM} provides an accurate fit,
matched only by the empirical fitting function of the Burkert profile,
while NFW is not favoured by the data.}
\label{fig:FornaxFitting}
\end{figure*}

For more massive galaxies, the solitons we predict are denser and more
massive, scaling approximately as $M_s \appropto M_{gal}^{1/3}$. So
for the Milky Way (MW), adopting a total mass of
$M_{gal}=10^{12}~\msun$, we expect a soliton of $M_s\simeq 2\times
10^9~\msun$, with core radius $\simeq 180~\pc$ and a potential depth
corresponding to a line-of-sight velocity dispersion
$\sigma_{||}\simeq 115~\rm{km/s}$ for test particles satisfying the
virial condition with the soliton potential. At face value this seems
consistent with the MW bulge velocity dispersion where a distinctive
flat peak is observed at a level of $\sigma_{||}\simeq 110~{\rm km/s}$
within a projected radius $\sim
200~\pc$\cite{Minniti1996,Rich2007,Ness2013}. Such cores clearly have
implications for the creation of spheroids acting as an essential
seed for the prompt attraction of gas within a deepened
potential. Indeed, bulge stars with [Fe/H] $>-1.0$ are firmly
established as a uniformly old population that formed
rapidly\cite{Zoccali2003,Ness2013}, a conclusion that standard
$\Lambda$CDM struggles to explain via extended accretion and
merging\cite{Ness2013}. The implications for early
spheroid formation and compact nuclear objects in general can be
explored self-consistently with the addition of baryons to the
{\psiDM} code, to model the interplay among stars, gas and {\psiDM}
that will provide model rotation curves for an important test of this
model.

At high redshift, the earliest galaxies formed from {\psiDM} are delayed
relative to standard CDM, limited by the small amplitude of Jeans mass
at radiation-matter equality, after which the first structures grow.
This is demonstrated with a {\psiDM} simulation of a $30~{\rm
h^{-1}}\Mpc$ box where we adopt $m_B=8.1\times 10^{-23}~\eV$ derived
above. The first bound object collapses at $z\simeq 13$, with a
clear solitonic core of mass $\simeq 10^9~\msun$ and radius
$\simeq 300~\pc$, whereas under $\Lambda$CDM the first objects
should form at $z\simeq 50$ with masses of only
$10^4-10^5~\msun$\cite{Abel2002}. The currently highest redshift galaxy at
$z\simeq 10.7$ is multiply lensed, appearing smooth and spherical,
with a stellar radius $\simeq 100~\pc$\cite{Coe2013}, similar to local
dSph galaxies. Deeper cluster lensing data from the Hubble ``Frontier
Fields'' programme will soon meaningfully explore the mass limits of
galaxy formation to higher redshift, allowing us to better distinguish
between particle and wavelike cold dark matter.

\bibliography{Letter}

\vspace*{-0.5cm}\paragraph*{Acknowledgements}
We thank Tak-Pong Woo for calculating
the soliton solution and Ming-Hsuan Liao for helping conduct the simulations.
The GPU cluster donated by Chipbond Technology Corporation, with which this
work is conducted, is acknowledged.
This work is supported in part by the National Science Council of Taiwan
under the grants NSC100-2112-M-002-018-MY3 and NSC99-2112-M-002-009-MY3.

\baselineskip14pt 
\setlength{\parskip}{4pt}
\setlength{\parindent}{18pt}%
\setlength{\footskip}{25pt}
\setlength{\textheight}{670pt}
\setlength{\oddsidemargin}{-8pt}
\setlength{\topmargin}{-41pt}
\setlength{\headsep}{18pt}
\setlength{\textwidth}{469pt}
\setlength{\marginparwidth}{42pt}
\setlength{\marginparpush}{5pt}

\addtolength{\topmargin}{-0.6cm}

\twocolumn

\setcounter{figure}{0}
\renewcommand{\figurename}{\textbf{Figure}}
\renewcommand{\thefigure}{\textbf{S\arabic{figure}}}

\twocolumn[\begin{@twocolumnfalse}

\vspace*{1cm}\centering{\textbf{{\Large Supplementary Information}}}

\vspace*{1cm}\textsf{\textbf{{\small
Hsi-Yu Schive$^{1}$,
Tzihong Chiueh$^{1,2\ast}$ \&
Tom Broadhurst$^{3,4}$}}}

\vspace*{0.5cm}\textbf{\textit{{\normalsize
chiuehth@phys.ntu.edu.tw}}}\vspace{1.8cm}

\end{@twocolumnfalse}]

\footnotetext[1]{\footnotesize Dept. of Physics, National Taiwan Univ.,
                               Taipei 10617, Taiwan}
\footnotetext[2]{\footnotesize Center for Theoretical Sciences, National
                               Taiwan Univ., Taipei 10617, Taiwan}
\footnotetext[3]{\footnotesize Dept. of Theoretical Physics, Univ. of the
                               Basque Country UPV/EHU, E-48080 Bilbao, Spain}
\footnotetext[4]{\footnotesize Ikerbasque, Basque Foundation for Science,
                               E-48011 Bilbao, Spain}

{\bf\small
This document provides supplementary information for the letter. We
begin by summarizing our simulation method, then we describe the soliton
solution of the Schr\"{o}dinger-Poisson equation and finally, we test
the soliton solution against the observed internal stellar dynamics of
dSph galaxies and constrain the boson mass $m_B$ of
the {\psiDM} model.
\vspace*{-0.3cm}}

\small

\vspace*{0.2cm}
\subsection*{Simulation method\vspace*{-0.2cm}}

The governing equation in the {\psiDM} model is the Schr\"{o}dinger-Poisson (SP)
equation, expressed here in the comoving coordinates;
\begin{equation}
\left[i\frac{\partial}{\partial\tau}+\frac{\nabla^2}{2}-aV\right]\psi=0
\label{eq:Schrodinger}
\end{equation}
and
\begin{equation}
\nabla^2 V= |\psi|^2-1,
\label{eq:Poisson}
\end{equation}
where $a$ is the cosmic scale factor and $V$ is the gravitational potential.
The comoving coordinates are normalized to the length
$\bm{\xi}\equiv(\frac{3}{2}H_0^2\Omega_{m0})^{1/4}(m_B/\hbar)^{1/2}{\bm{x}}$
and the normalized timestep
$d\tau\equiv (\frac{3}{2}H_0^2\Omega_{m0})^{1/2}a^{-2}dt$, where $H_0$ is 
the present Hubble parameter, $\Omega_{m0}$ the present dark matter density
parameter, and $m_B$ the particle mass. The comoving mass density
$\rho=|\psi|^2$ is normalized to the comoving background density $\left<\rho\right>$.

In earlier work we adopted a pseudospectral method to
simulate the {\psiDM} model with a uniform mesh resolution\cite{WC2009}, but the spatial
resolution achieved ($1024^3$ grid) was inadequate for the
innermost regions of the haloes. To form a minimum of a few tens
of dSph mass objects, the simulation volume must span
a few cubic $\Mpc$. On the other hand, to be capable of resolving the compact
cores with sizes of few hundreds $\pc$ that we find forming within each halo,
the spatial resolution must achieve at least few tens $\pc$. Hence the dynamical range in scale
is $\sim 10^5$, which is infeasible with a uniform mesh.

To solve this issue, we have developed a highly optimized AMR framework,
GAMER\cite{GAMER2010,Schive2012}, featuring an extremely efficient solution to 
integrating the AMR method with graphic processing units (GPUs) for computation acceleration. We
incorporate an octree AMR algorithm\cite{PARAMESH2000}, and use CPUs to manipulate the AMR data
structure and GPUs to accelerate the partial differential equation solvers
for the SP equation. To optimize the simulation performance, we implement
the asynchronous data transfer between CPUs and GPUs and the hybrid
MPI/OpenMP/GPUs parallelization, and fully exploit the simultaneity of
CPUs' and GPUs' computation. The workload balance among multiple GPUs is achieved by the
Hilbert space-filling curve method. As a result, an overall performance speedup
up to 40 has been demonstrated in various GPU clusters for the {\psiDM}
simulations.

We advance the wave function
forward in time by applying a unitary transformation
$\psi(\tau+\triangle \tau)=e^{-iW\triangle \tau}e^{-iK\triangle \tau}\psi(\tau)$, 
which is an approximate solution to the SP equation
when the evolution timestep $\triangle \tau$ is small.
Here $K~(\equiv-\nabla^2/2)$ and $W~(\equiv aV)$ are the kinematic and
potential energy operators, respectively. The unitary operator
$e^{-iK\triangle \tau}$ is expanded explicitly to order
$\triangle \tau^{N_K}$. From the von Neumann stability
analysis\cite{NumericalRecipes} we find that $N_K \le 2$ is unconditionally 
unstable. In practice, we choose $N_K=5$ with
modified Taylor expansion coefficients in order to minimize the
small-scale numerical damping. The gravitational potential is calculated via a
multi-level Poisson solver, where the Fast Fourier Transform
(FFT)\cite{FFTW2005} method with periodic boundary condition is applied at
the AMR root level (the lowest resolution level covering the
entire simulation volume). Either the successive overrelaxation (SOR)
or the multigrid Poisson solver\cite{NumericalRecipes} is adopted at the
refinement levels. Second-order accuracy has been verified in a
variety of tests. Technicalities of the numerical schemes are to be
detailed in a separate paper (Schive et al., in preparation).

The evolution timesteps are determined by $d\tau\le
min[4C_K\triangle\xi^2/(3\pi),~2\pi C_W / (a|V|_{max})]$, where the
two terms in the square bracket stem from the stability consideration
of $K$ and $W$ operators, respectively. Here
$C_K~(=0.625)$ and $C_W~(=0.3)$ are parameters controlling the
integration accuracy. Having $C_K\le1.0$ and $C_W\le1.0$ ensures that
the phase angle of wave function rotated in one timestep is smaller
than $2\pi$. Note that the kinematic solver requires
$d\tau\propto\triangle\xi^2$, a signature of the $\nabla^2$
operator. This unpleasant scaling leads to an extremely small timestep
($\triangle a \sim 1\times10^{-6}$) when high spatial resolution
($\sim10^2~\pc$) is required. To alleviate this issue, we have adopted
the individual timestep scheme, with which lower resolution regions
are allowed to have larger timesteps and the smallest timestep is
applied only to the highest resolution regions, normally occupying
less than 1\% of the entire simulation volume.

Upon simulating the wave mechanics, the grid refinement criteria need
to be carefully designed in order to achieve appropriate resolution
everywhere. In particular, since the flow velocity can be expressed
as $\bm{v}\equiv\nabla S$, where we let $\psi \equiv f e^{iS}$, the
wave function will exhibit strong and rapid oscillation
in the regions with high velocity. Therefore, unlike the conventional
CDM simulations with AMR, where in general higher spatial resolution is
required only around the regions with higher density or density
contrast, here we also need sufficient spatial and temporal
resolution to resolve the high-speed flow, even if the density is
low and smooth (Fig. S1). However, using flow velocity $\bm{v}$ directly as the
refinement criterion is impractical since $\bm{v}$ is not an
observable and may also diverge (e.g., in the locations of
zero density resulting from interference). To solve the issue, we
have applied the L\"{o}hner's error estimator\cite{Lohner1987}, which basically
estimates the ratio between the second and first derivatives of both
the real and imaginary parts of the wave function. Specifically, we demand
there should be at least four cells to resolve one wavelength. In
addition, we prohibit grid refinement in regions of extremely
low density, where quantized vortices can form\cite{Chiueh1998,Chiueh2011} but are not
relevant to this work. Finally, any cell with enclosed mass larger
than $1.5\times10^5~\msun$ (which is equal to
$8\left<\rho\right>\triangle\xi_0^3$, where $\triangle\xi_0$ is the root-level
cell size) is forced to be refined so as to capture
the core structure.

\begin{figure*}[t]
\centering
\includegraphics[width=7.9cm,clip=true,trim=0cm 0cm 0cm 0cm]{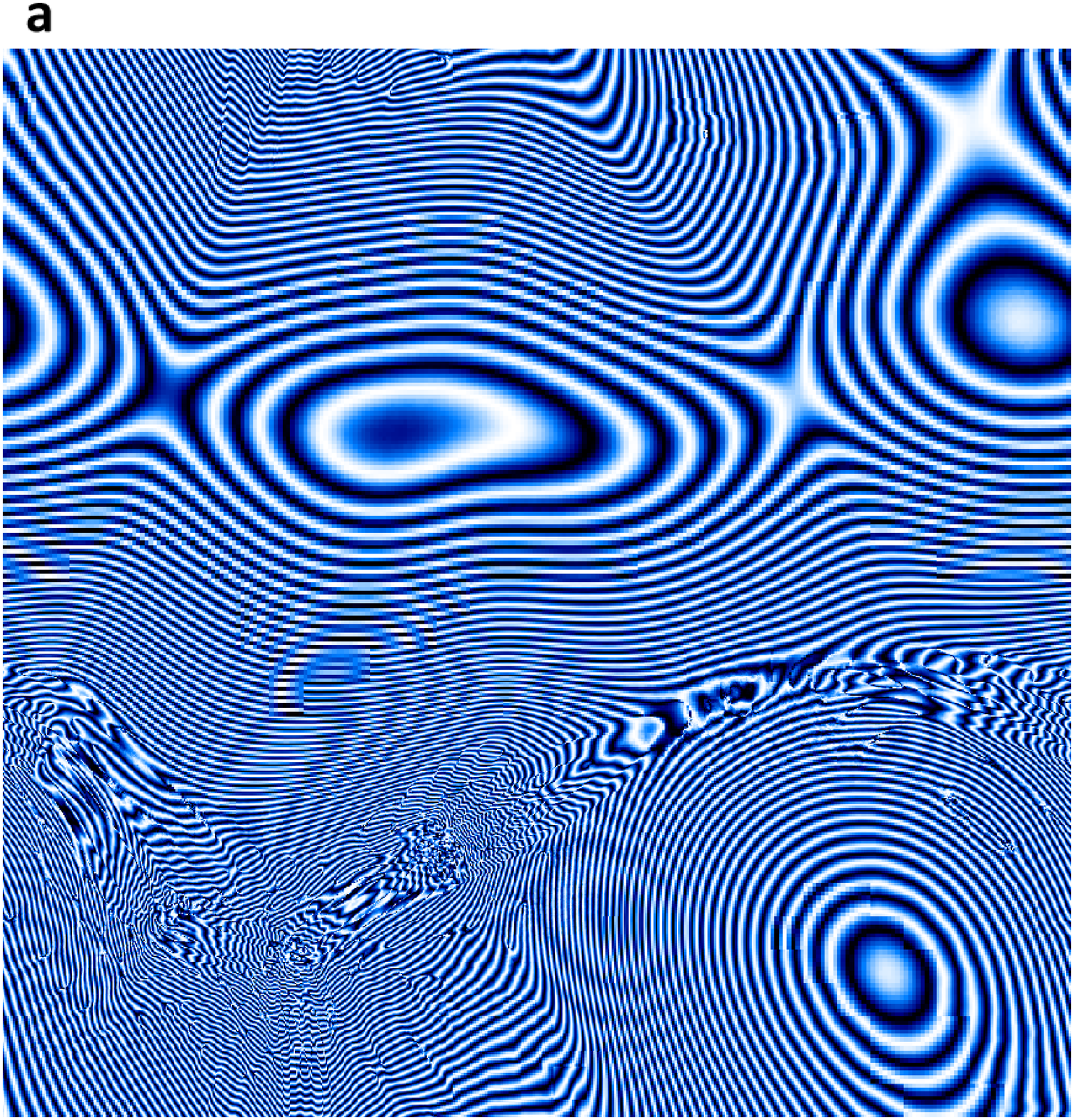}
\includegraphics[width=7.9cm,clip=true,trim=0cm 0cm 0cm 0cm]{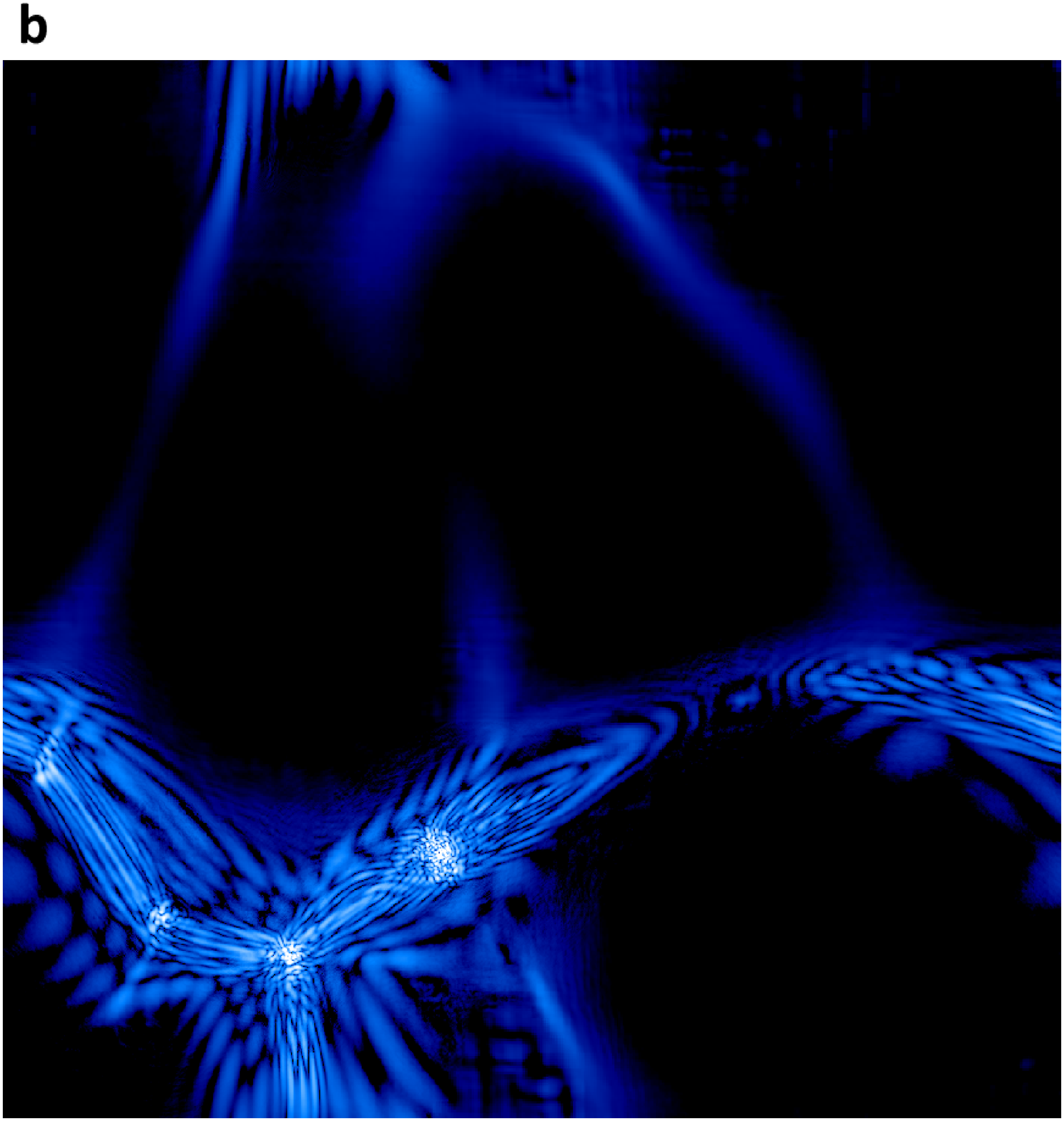}
\caption{\textbf{Square wave function $\bm{\psi^2~(\equiv f^2e^{i2S})}$ in the {\psiDM} simulation.}
Panels (\textbf{a}) and (\textbf{b}) show a $2~\Mpc$ slice of phase ($sin(2S)$) and
amplitude ($f^2$) of the wave function at $z=3.1$, respectively. 
The simulation challenge arises from the complexity of the wave function.
Strong and rapid phase oscillations are common everywhere (even in the 
low-density background shown by the dark regions in the density plot), 
where sufficient spatial and temporal resolution is required to resolve 
each wavelength.}
\label{fig:WaveFunction}
\end{figure*}

The volume of our detailed simulation is $2~\Mpc$ on a side, with a base-level grid
$N=256^3$ and up to seven refinement levels, giving an effective
resolution $60~\pc$. The initial linear power spectrum was
constructed by CMBFAST\cite{CMBFAST} in a $\Lambda$CDM universe at
$z=1,000$, with cosmological parameters consistent with the recent
observation\cite{WMAP9}. The relatively high initial redshift adopted here
compared with traditional CDM simulations ensures substantial high-k
damping of $\sim 10^3$ in the linear power spectrum when reaching
$z\sim30$ (Fig. S2).
Note that in our comparison of {\psiDM} and particle CDM evolution, shown in
Fig. 1, we scale the {\psiDM} linear power spectrum at $z=30$ back to
$z=100$ as the initial condition for particle CDM simulation in order
to highlight the large-scale features.

\begin{figure}[t]
\centering
\vspace*{0.2cm}
\includegraphics[width=7.9cm]{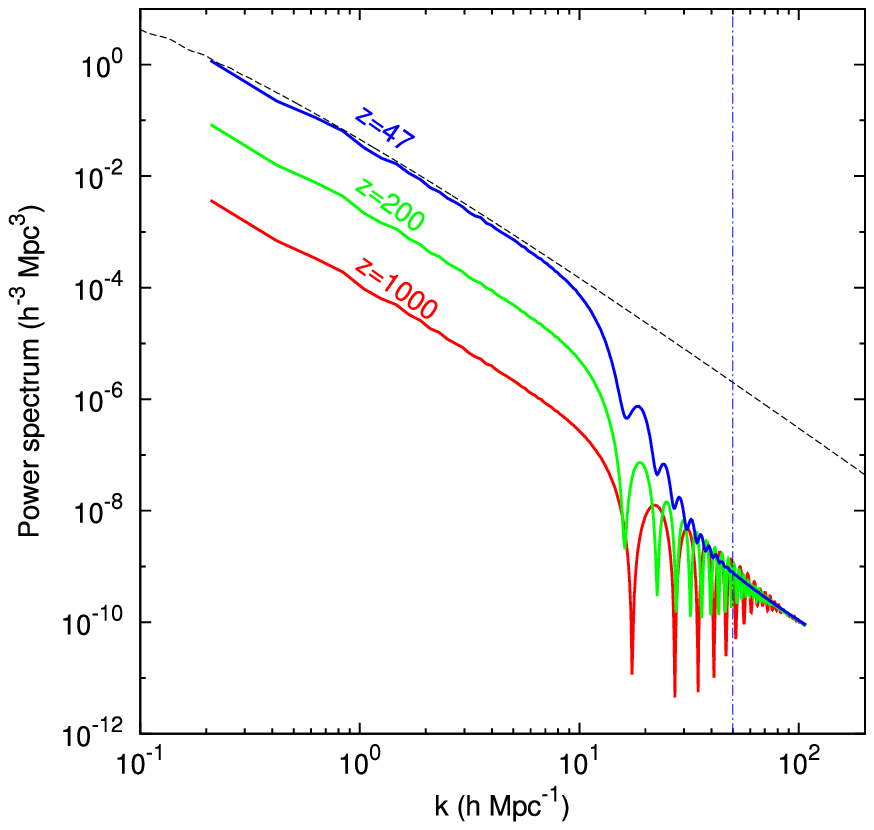}
\caption{\textbf{Dark matter power spectrum at various epochs.}
Solid lines show the power spectra obtained for the {\psiDM} model with a $30~{\rm h^{-1}}\Mpc$
simulation box and $1024^3$ resolution, and the vertical dot-dashed line shows
the Jeans length at $z=47$. The dashed line denotes the predicted
linear power spectrum in the conventional CDM model at $z=47$.}
\label{fig:LinearPowerSpectrum}
\end{figure}

\subsection*{Soliton solution\vspace*{-0.2cm}}

A soliton solution to the SP equation can be found
numerically as follows. Firstly, when
deriving the soliton profile, it is reasonable to assume $a=1$ and
$|\psi|^2 \gg 1$ in Eqs. (\ref{eq:Schrodinger}) and (\ref{eq:Poisson})
even in a cosmological context since the characteristic wave crossing time
around the core is much shorter than the Hubble time and the
core density is at least several orders of magnitude higher than the
background density. Then, by assuming spherical symmetry and
inserting the stationary condition
$\psi(\xi,\tau)=e^{-i\omega\tau}\Psi(\xi)$, the dimensionless SP
equation can be further reduced to a coupled second-order ordinary
differential equation that can be solved numerically with proper
boundary conditions\cite{Guzman2006}. The
soliton profile is close to Gaussian, with a near constant-density
core and a steeper outer gradient (Fig. S3). We define a core radius $r_c$
at which the density has dropped to one-half its peak value. The
corresponding core mass $M_c\equiv M(r\le r_c)$ encloses roughly $1/4$
of the total soliton mass $M_s\equiv M(r\to\infty)$. The half-mass
radius is $\sim 1.45~r_c$.

\begin{figure}[t]
\centering
\vspace*{0.2cm}
\includegraphics[width=7.9cm]{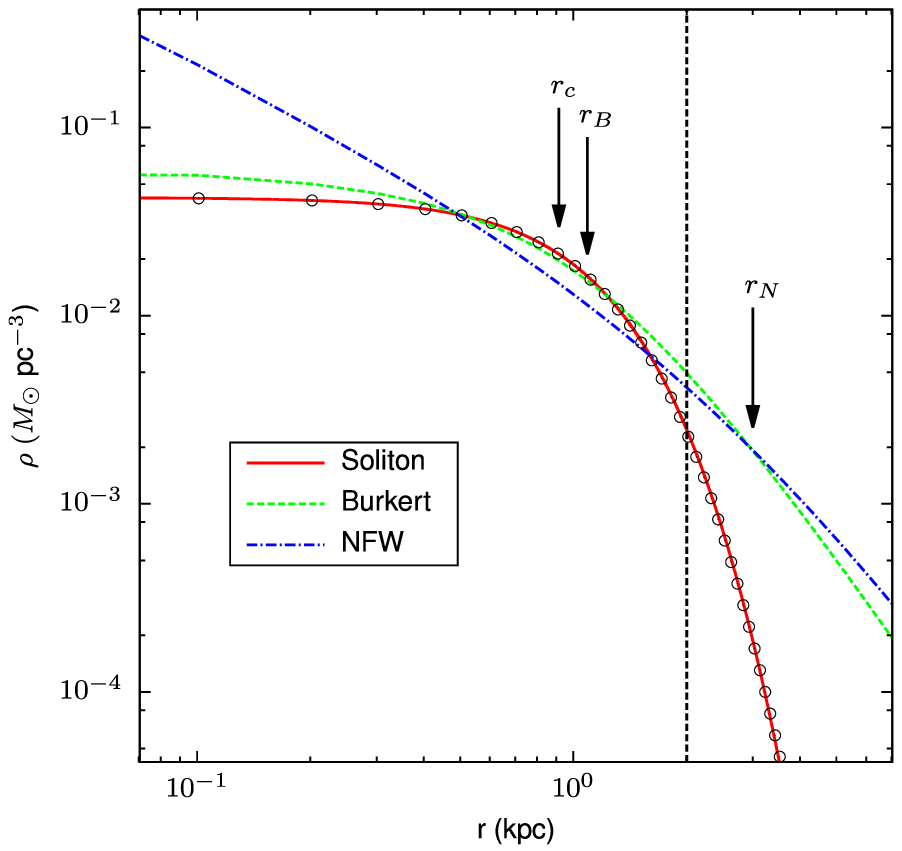}
\caption{\textbf{Soliton density profile.}
The red solid line shows the numerical
solution, and circles show the approximate analytical form
(Eq. [\ref{eq:Soliton}]) found to fit well to the soliton profile within
$3\,r_c$. For comparison, we also show the empirically motivated Burkert (green dashed line) and
NFW (blue dot-dashed line) profiles. All three model curves are plotted
using the best-fit parameters to Fornax, and the vertical dashed line represents
the upper limit of radius adopted for the fitting. Arrows indicate the soliton
core radius ($r_c$), the scale radius of Burkert ($r_B$), and NFW ($r_N$),
respectively.}
\label{fig:SolitonProfile}
\end{figure}

An important feature of the soliton solution to appreciate is its
scaling symmetry\cite{Guzman2006}. The wave function and the associated physical
quantities allow for a scale transformation, $(r,\psi,\rho_s,M_s)\to
(\lambda^{-1}r,\lambda^2\psi,\lambda^4\rho_s,\lambda M_s)$, where
$\rho_s$ is the soliton density profile, to generate other solutions,
thus forming a one-parameter family. Accordingly, all soliton
solutions can be characterised by a single parameter (for example,
$r_c$), providing clear predictions for the correlation between
different core properties. For example, if the core radius of a galaxy
is observed to be half the size of another, the soliton
solution predicts the core mass and peak density to be two and
sixteen times higher.

The soliton profile does not have an analytical form and the solution can
only be obtained numerically. But thanks to the scaling symmetry,
the core mass and core radius obey a simple relation
\begin{equation}
M_c\approx\frac{5.5\times10^9}{(m_B/10^{-23}~\eV)^2(r_c/\kpc)}~\msun.
\label{eq:SolitonMass}
\end{equation}
For example, with the best-fit for Fornax of $m_B=8.1\times10^{-23}~\eV$ and
$r_c=0.92~\kpc$, we readily have
$M_c\sim9.1\times10^7~M_\odot$ and $M_s\sim3.6\times10^8~M_\odot$. In addition, it is found that within the
range $0\le r\lesssim 3\,r_c$, which encloses $\sim 95\%$ of the total soliton
mass, the soliton density profile can be well approximated by
\begin{equation}
\rho_s(r)\approx\frac{1.9~(m_B/10^{-23}~\eV)^{-2}(r_c/\kpc)^{-4}}{[1+9.1\times10^{-2}(r/r_c)^2]^8}~\msun\pc^{-3},
\label{eq:Soliton}
\end{equation}
which is consistent with the scaling relation that the peak density is
proportional to $r_c^{-4}$ for a given particle mass. This approximate
analytical formula makes it convenient to compare the
soliton model and observation, from which the best-fit $m_B$ and $r_c$ can
be determined (see the next section).

\subsection*{Data modeling\vspace*{-0.2cm}}

The dSph galaxies are the most dark-matter dominated objects known,
as indicated from their high mass-to-light ratios and hence
very useful for determining the properties of dark matter on
small scales. By assuming spherical symmetry and dynamical
equilibrium, the total mass profile $M(r)$ can be related to the observed
distribution of stars and their velocity dispersion profile
via the Jeans equation\cite{GalacticDynamics2008}:
\begin{equation}
\frac{d(\rho_\star\sigma_r^2)}{dr}=-\rho_\star\frac{d\Phi}{dr} - \frac{2\beta\rho_\star\sigma_r^2}{r},
\label{eq:Jeans}
\end{equation}
where $\rho_\star(r)$ is the stellar density, $\sigma_r(r)$ describes the
radial velocity dispersion, $\beta(r)$ quantifies the stellar velocity
anisotropy, and $\Phi(r)$ is the gravitational potential satisfying
$d\Phi/dr=GM(r)/r^2$. For the observed line-of-sight velocity dispersion
profile and the projected stellar distribution, one can then parameterise
$\rho_\star(r)$, $\beta(r)$, $M(r)$ and determine the best fit to the data.
Unfortunately, none of the mass models adopted in previous work resemble the
soliton profile found in the {\psiDM} model.

We first examine the well studied Fornax dSph galaxy as a
benchmark. Fornax is found to have three stellar
subpopulations inferred from their different metallicities,
kinematics, and spatial distribution\cite{AE2012a}. Although these
subpopulations reside in the same gravitational potential, they have
different half-light radii and hence sample different volumes of the
core. The predominant intermediate metallicity subpopulation has a
projected half-light radius $R_h\sim 0.61~\kpc$ and a nearly constant
line-of-sight velocity dispersion $\sigma_{||}\sim11.3\pm0.7~{\rm
km/s}$ within $0\le r\lesssim 1.4~\kpc$, indicative of an
isotropic velocity dispersion
($\beta\sim0$). The simplicity of this subpopulation leads to an exact
solution to Eq. (\ref{eq:Jeans}), $\rho_\star(r)=\rho_0
\exp[-\phi(r)/\sigma_r^2]$, allowing us to determine the stellar
density beyond the half-light radius. By making a further assumption
that the stellar self-gravity is negligible, the gravitational
potential $\phi(r)$ of dark matter can be calculated analytically from
Eq. (\ref{eq:Soliton}) for a given particle mass and core radius,
which we show in Fig. S4 for a range of confidence levels.

\begin{figure}[t]
\centering
\vspace*{0.2cm}
\includegraphics[width=7.9cm]{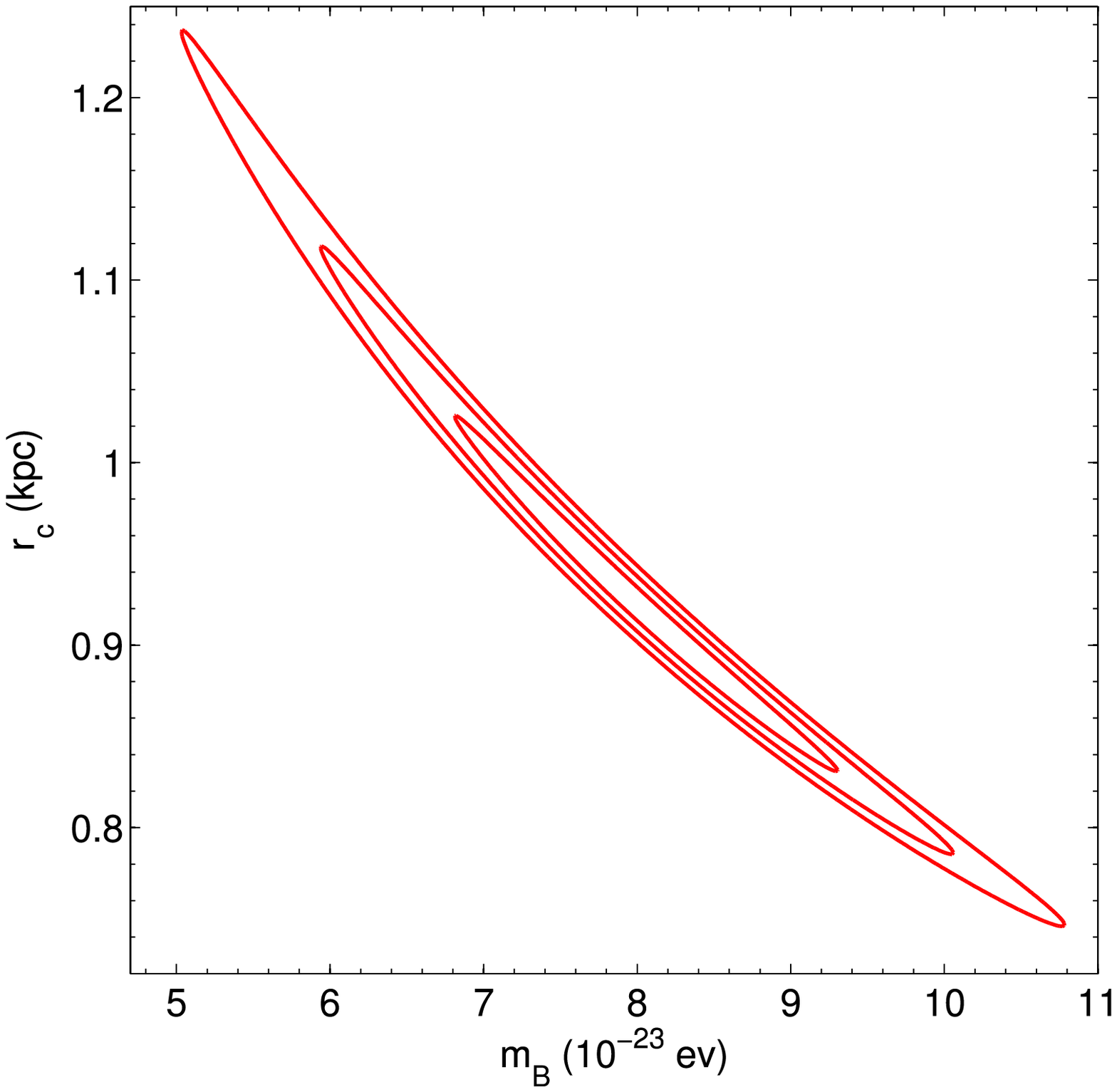}
\caption{\textbf{Confidence regions of the dark matter mass and core radius for the
soliton profile fit to Fornax.}
Contours show the regions of 68\%, 95\%, and 99.7\% confidence, respectively.}
\label{fig:ParticleMassContour}
\end{figure}

Notice the tight correlation between $m_B$ and $r_c$ shown in
Fig. S4, namely $m_B\propto r_c^{-1.5}$.
This results from the fact that nearly two-third of the stars used in this
fit are located inside the core radius, by which only the peak core
density is constrainable, with $m_B$ and $r_c$ to be
degenerate in this limit as $m_B\propto r_c^{-2}$ (referred to Eq. [\ref{eq:Soliton}]).
Nevertheless, one third of the stars still lie outside the core
radius so that the degeneracy between $m_B$ and $r_c$ is broken to some degree. By comparison,
the metal-rich subpopulation is more concentrated inside the core, from which
$m_B$ and $r_c$ will be poorly constrained individually.

In the following we describe several consistency checks
for the $m_B$ and $r_c$ determined above. Firstly, the three separate stellar
subpopulations of Fornax also provide three independent enclosed mass estimates
along the mass profile of Fornax\cite{WP2011,AAE2013}.
Each stellar subpopulation has a different half-light radius,
$R_h\sim430-940~\pc$, and velocity dispersion,
$\sigma_{||}\sim7-17~{\rm km/s}$, for which the enclosed mass can be related
to mean radius by a model-independent expression
$M(1.67\,R_h)\simeq5.85\,R_h\sigma_{||}^2(R_h)/G$ - a simple well tested 
relation\cite{Walker2010,AE2012b}. The consistency
between the soliton and Fornax mass profiles is verified within the
1-$\sigma$ confidence region for each point (see Fig. 4b), although the soliton mass profile is a bit
higher at $r\sim 1~\kpc$. It is worthwhile to note that the metal-poor
subpopulation has the largest mean radius, $R_h\sim935~\pc$, providing a mass estimate for $r\sim1.6~\kpc$
which is greatly beyond the core radius. This subpopulation
therefore provides a strong constraint on the density profile outside
the constant-density region and substantially breaks the $m_B-r_c$ degeneracy.

The existence of five old globular clusters in Fornax provides another
evidence for the large core\cite{Goerdt2006,Cole2012}.
These globular clusters reside at $\sim 1~\kpc$ in radius from the centre of 
Fornax and their dynamical friction timescale is determined from N-body
calculations to be far shorter than the Hubble time, so that these clusters
should be found at the centre of Fornax if the density profile of Fornax
followed the predicted cuspy form of particle CDM. 
Instead, for these objects to be dynamically stable at their observed radii,
a cored profile is indicated, of a similar radius
again consistent with our predicted core radius for Fornax.

Stars in the Sculptor dSph have also been reliably separated by metallicity into metal-rich
and metal-poor subpopulations\cite{WP2011,AE2012b}. Enclosed masses of the two subpopulations
have been determined to be $M(r\lesssim167~\pc)\sim4.1\times10^6~\msun$ and
$M(r\lesssim302~\pc)\sim2.4\times10^7~\msun$, respectively\cite{WP2011}. It is
found that $M(r)\propto r^3$, indicative of both subpopulations residing well
within a flat core, thus preventing us from constraining $m_B$ and $r_c$
separately. On the other hand, by comparing
with the mass density of Fornax and using the soliton density scaling
$\rho(r=0)\propto r_c^{-4}$, we estimate the core size of Sculptor to be
$\sim 610~\pc$, verifying that the two subpopulations are indeed well 
within the core.

A third dSph galaxy that favors the cored dark matter profile in the literature
is Ursa Minor\cite{Kleyna2003,Lora2012}, which is one of the most dark-matter dominated classical
Milky Way dwarf satellites, with a mass-to-light ratio
$\sim70~\msun/L_\odot$. A kinematically cold, and dense old star cluster
is found in Ursa Minor, and it was shown by N-body simulations that such a
cold clump can be easily destroyed within less than 1 Gyr in a cuspy
density profile, hence incompatible with the standard CDM model. 
By comparison, it was demonstrated that the cold substructure can be stable for a Hubble time
if residing in a harmonic dark matter potential of a flat cored profile with
$r_c\gtrsim 450~\pc$. To check whether it accords with our
best-fit soliton model, we take $M(r<280~\pc)=1.3\times10^7~\msun$\cite{Walker2010}
and assume this mass to be well within the core, from which we estimate
$r_c\sim 680~\pc$, consistent with the core size estimated by other studies.

\bibliography{Supplement}

\end{document}